\documentclass[sigconf,9pt,compact]{acmart}
\pdfoutput=1

\renewcommand\footnotetextcopyrightpermission[1]{} % removes footnote with conference information in first column
\usepackage{paralist}
\usepackage[listings,skins]{tcolorbox}

\long\def\diffcolor#1#2\@nil{color\string#1diff}

\def\verbatim@processline{%
\nointerlineskip\noindent\rlap{%
\colorbox{\expandafter\diffcolor\next..\@nil}{%https://www.overleaf.com/project/5dd647a01f3952000193d6c2
\the\verbatim@line}}\par}
\makeatother
\definecolor{diffstart}{rgb}{0.7, 0.75, 0.71}
\definecolor{diffincl}{rgb}{0.0, 0.42, 0.24}
\definecolor{diffrem}{rgb}{1.0,0.44, 0.37}

\AtBeginDocument{%
  \providecommand\BibTeX{{%
    \normalfont B\kern-0.5em{\scshape i\kern-0.25em b}\kern-0.8em\TeX}}}

\begin{document}
\settopmatter{printfolios=true,authorsperrow=4}
\title{Superoptimization of WebAssembly Bytecode}

%\author{
%Javier Cabrera Arteaga
%\texttt{javierca@kth.se}
%\and
%Shrinish Donde
%}

\author{Javier Cabrera Arteaga}
\affiliation{}
\email{javierca@kth.se}

\author{Shrinish Donde}
\affiliation{}
\email{shrinish@kth.se}

\author{Jian Gu}
\affiliation{}
\email{jiagu@kth.se}

\author{Orestis Floros}
\affiliation{}
\email{forestis@kth.se}

\author{Lucas Satabin}
\affiliation{}
\email{lucas.satabin@gnieh.org}

\author{Benoit Baudry}
\affiliation{}
\email{baudry@kth.se}

\author{Martin Monperrus}
\affiliation{}
\email{martin.monperrus@csc.kth.se}

%% \author{G.K.M. Tobin}
%% \authornotemark[1]
%% \email{webmaster@marysville-ohio.com}
%% \affiliation{%
%% \institution{Institute for Clarity in Documentation}
\newcommand{\red}[1]{{\color{red} #1 }}
\sloppy
%% \renewcommand{\shortauthors}{Trovato and Tobin, et al.}

\begin{abstract}
Motivated by the fast adoption of WebAssembly, we propose the first functional pipeline to support the superoptimization of WebAssembly bytecode. Our pipeline works over LLVM and Souper. We evaluate our superoptimization pipeline with 12 programs from the Rosetta code project. Our pipeline improves the code section size of 8 out of 12 programs. We discuss the challenges faced in superoptimization of WebAssembly with two case studies.
\end{abstract}
% \keywords{Superoptimization, WebAssembly, Web, Optimization, Semantic equivalence, WASM programs, LLVM IR}
\maketitle

\section{Introduction}

% Intro to WASM
After HTML, CSS, and JavaScript, WebAssembly (WASM) has become the fourth standard language for web development \cite{W3C}. This new language has been designed to be fast, platform-independent, and experiments have shown that WebAssembly can have an overhead as low as 10\% compared to native code \cite{haas_bringing_nodate}.
Notably, WebAssembly is developed as a collaboration between vendors and has been supported in all major browsers since 2017 \cite{WebAssembly2016}. 

%Today, it is perceived as a promising platform-independent binary format \todo{\cite{X} }.

% Motivation
The state-of-art compilation frameworks for WASM are Emscripten and LLVM \cite{noauthor_emscripten-core/emscripten_2019, llvm}, they generate WASM bytecode from high-level languages (e.g. C, C++, Rust). These frameworks can apply a sequence of optimization passes to deliver smaller and faster binaries. In the web context, having smaller binaries is important, because they are delivered to the clients over the network, hence smaller binaries means reduced latency and page load time. Having smaller WASM binaries to reduce the web experience is the core motivation of this paper.

% intro superoptimization
To reach this goal, we propose to use superoptimization.
Superoptimization consists of synthesizing code replacements in order to further improve binaries, typically in a way better than the best optimized output from standard compilers \cite{Sasnauskas2017a, churchill_sound_nodate}.
Given a program, superoptimization searches for alternate and semantically equivalent programs with fewer instructions \cite{Massalin1987}. In this paper, we consider the superoptimization problem stated as finding an equivalent WebAssembly binary such that the size of the binary code is reduced compared to the default one.

% Contribution
This paper presents a study on the feasibility of superoptimization of WebAssembly bytecode.
We have designed a pipeline for WASM superoptimization, done by tailoring and integrating open-source tools.
Our work is evaluated by building a benchmark of 12 programs and applying superoptimization on them. The pipeline achieves a median size reduction of 0.33\% in the total number of WASM instructions.

To summarize, our contributions are:
\begin{compactitem}
    \item The design and implementation of a functional pipeline for the superoptimization of WASM.
    \item Original experimental results on superoptimizing 12 C programs from the Rosetta Code corpus.
\end{compactitem}

\section{Background}

\subsection{WebAssembly}\label{wasm}

% What is WASM
WebAssembly is a binary instruction format for a stack-based virtual machine \cite{WebAssembly2016}. As described in the WebAssembly Core Specification \cite{W3C}, WebAssembly is a portable, low-level code format designed for efficient execution and compact representation. WebAssembly has been first announced publicly in 2015. Since 2017, it has been implemented by four major web browsers (Chrome, Edge, Firefox, and Safari). A paper by \citet{haas_bringing_nodate} formalizes the language and its type system, and explains the design rationale.

% https://webassembly.github.io/spec/core/binary/modules.html#sections
% code, data, imports, exports, types definition, memory, start definition, tables and elements.

The main goal of WebAssembly is to enable high performance applications on the web.
WebAssembly can run as a standalone VM or in other environments such as Arduino \cite{WARDuino2019}.
It is independent of any specific hardware or languages and can be compiled for modern architectures or devices, from a wide variety of high-level languages.
In addition, WebAssembly introduces a memory-safe, sand-boxed execution environment to prevent common security issues, such as data corruption and security breaches.

%\subsection{WebAssembly Program}
%The WebAssembly Core Specification \cite{WebAssembly2016_spec} defines WebAssembly as a programming language expressed either in binary or textual format, and both are of a common structure. A WASM program is designed to be a separate module containing collection of various wasm-defined values and types.

Since version 8, the LLVM compiler framework supports the WebAssembly compilation target by default \cite{llvm}. This means that all languages that have an LLVM front end can be directly compiled to WebAssembly. Binaryen \cite{Binaryen}, a compiler and toolchain infrastructure library for WebAssembly, supports compilation to WebAssembly as well. Once compiled, WASM programs can run within a web browser or in a standalone runtime \cite{WARDuino2019}.

\vspace{-6mm}
\subsection{Superoptimization}\label{SO}

Given an input program, code superoptimization focuses on \emph{searching} for a new program variant which is faster or smaller than the original code, while preserving its correctness \cite{bunel_learning_2017}.
The concept of superoptimizing a program dates back to 1987, with the seminal work of Massalin \cite{Massalin1987} which proposes an exhaustive exploration of the solution space. The search space is defined by choosing a subset of the machine's instruction set and generating combinations of optimized programs, sorted by length in ascending order. If any of these programs are found to perform the same function as the source program, the search halts. However, for larger instruction sets, the exhaustive exploration approach becomes virtually impossible.
Because of this, the paper proposes a pruning method over the search space and a fast probabilistic test to check programs equivalence.

% Massalin proposes two fixes to deal with the search time, a fast probabilistic test to check programs equivalence and a pruning method over the search space. This work then can be summarized in an optimization problem with two main components, the programs equivalence and the new program cost.
% This method works well for small instruction sets. 

State of the art superoptimizers such as STOKE~\cite{schkufza_stochastic_2013} and Souper~\cite{Sasnauskas2017a} make modifications to the code and generate code rewrites.
A cost function evaluates the correctness and performance of the rewrites.
Correctness is generally estimated by running the code against test cases (either provided by the user or generated automatically, e.g. symbolic evaluation on both original and replacement code). 

\subsection{Souper}
\label{souper}
Souper is a superoptimizer for LLVM \cite{Sasnauskas2017a}. It enumerates a set of several optimization candidates to be replaced. An example of such a replacement is the following, replacing two instructions by a constant value:

\begin{lstlisting}[language=C++,basicstyle=\small\ttfamily]
%0:i32 = var (range=[1,0))
%1:i1 = ne 0:i32, %0
\end{lstlisting}
\vspace{-.3cm}
\rule{\linewidth}{0.5pt}
\vspace{-.6cm}
\begin{lstlisting}[language=C++,basicstyle=\small\ttfamily]
cand %1 1:i1
\end{lstlisting}

In this case, Souper finds the replacement for the variable \texttt{\%1} as a constant value (in the bottom part of the listing) instead of the two instructions above.

Souper is based on a Satisfiability Modulo Theories (SMT) solver. SMT solvers are useful for both verification and synthesis of programs \cite{10.1007/978-3-540-78800-3_24}. With the emergence of fast and reliable solvers, program alternatives can be efficiently checked, replacing the probabilistic test of Massalin \cite{Massalin1987} as mentioned in \autoref{SO}.

In the code to be optimized, Souper refers to the optimization candidates as \emph{left-hand side} (LHS). Each LHS is a fragment of code that returns an integer and is a target for optimization. Two different LHS candidates may overlap. For each candidate, Souper tries to find a \emph{right-hand side} (RHS), which is a fragment of code that is combined with the LHS to generate a replacement. In the original paper's benchmarks \cite{Sasnauskas2017a}, Souper optimization passes were found to further improve the top level compiler optimizations (-O3 for \textit{clang}, for example) for some programs. 

Souper is a platform-independent superoptimizer. The cost function is evaluated on an intermediate representation and not on the code generated for the final platform. Thus, the tool may miss optimizations that make sense for the target instruction set.
%In this paper, we elaborate on this work and show how to use Souper for superoptimizing WebAssembly code.

\section{WASM Superoptimization Pipeline}
\label{pipeline}
\begin{figure}
\centering
\includegraphics[width=0.85\linewidth]{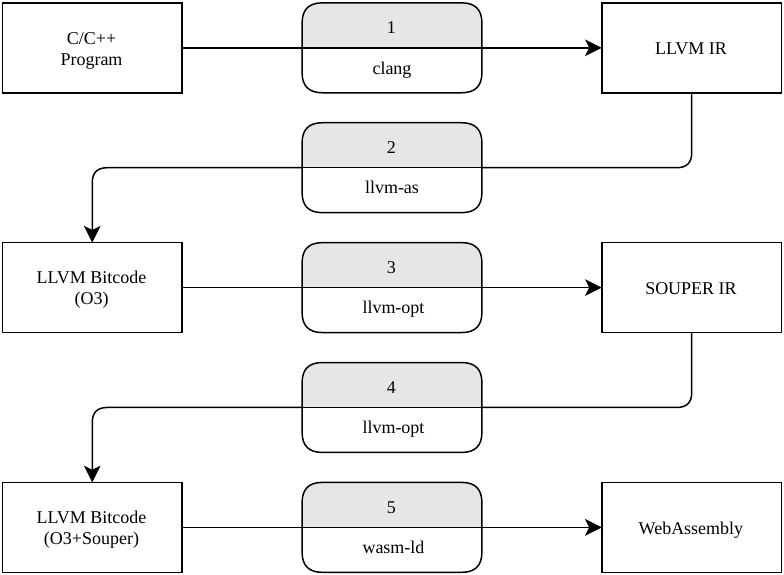}
\caption{Superoptimization pipeline for WebAssembly based on Souper}
\label{pipe}
\end{figure}

The key contribution of our work is a superoptimization pipeline for WebAssembly. We faced two challenges while developing this pipeline: the need for a correct WASM generator, and the usage of a full-fledged superoptimizer. The combination of the LLVM WebAssembly backend and Souper provides the solution to tackle both challenges.

\subsection{Steps}

Our pipeline is a tool designed to output a superoptimized WebAssembly binary file for a given C/C++ program that can be compiled to WASM. With our pipeline, users write a high level source program and get a superoptimized WebAssembly version. 

The pipeline (illustrated in \autoref{pipe}) first converts a high-level source language (e.g. C/C++) to the LLVM intermediate representation (LLVM IR) using the Clang compiler (Step 1). We use the code generation options in clang in particular the -O3 level of optimization which enables aggressive optimizations. In this step, we make use of the LLVM compilation target for WebAssembly `wasm32-unknown-unknown'. This flag can be read as follows:
wasm32 means that we target the 32 bits address space in WebAssembly; the second and third options set the compilation to any machine and performs inline optimizations with no specific strategy. LLVM IR is emitted as output.

% LLVM assembler
Secondly, we use the LLVM assembler tool (llvm-as) to convert the generated LLVM IR to the LLVM bitcode file (Step 2). This LLVM assembler reads the file containing LLVM IR language, translates it to LLVM bitcode, and writes the result into a file. Thus, we make use of the optimizations from clang and the LLVM support for WebAssembly before applying superoptimization to the generated code.
% optimization pass

Next, we use Souper, discussed in \autoref{souper}, to add further superoptimization passes. Step 3 generates a set of optimized candidates, where a candidate is a code fragment that can be optimized by Souper. From this, Souper carries out a search to get shorter instruction sequences and uses an SMT solver to test the semantic equivalence between the original code snippet and the optimized one \cite{Sasnauskas2017a}.

% opt tool
Step 4 produces a superoptimized LLVM bitcode file. The \textbf{opt} command is the LLVM analyzer that is shipped with recent LLVM versions. The purpose of the \textbf{opt} tool is to provide the capability of adding third party optimizations (plugins) to LLVM. It takes LLVM source files and the optimization library as inputs, runs the specified optimizations and outputs the optimized file or the analysis results. Souper is integrated as a specific pass for LLVM \textbf{opt}.

%At this point we make use of the WebAssembly support of LLVM to generate WebAssembly binaries with a tool called `wasm' \todo{wasm does not appear in the figure: all tools of the text must appear in the figure and vice versa}.
The last step of our pipeline consists of compiling the generated superoptimized LLVM bitcode file to a WASM program (Step~5). This final conversion is supported by the WebAssembly linker (wasm-ld) from the LLD project \cite{LLVM2019WebAssemblyDocumentation}. wasm-ld receives the object format (bitcode) that LLVM produces when run with the `wasm32-unknown-unknown' target and produces WASM bytecode. 

To our knowledge, this is the first successful integration of those tools into a working pipeline for superoptimizing WebAssembly code. 

\subsection{Insights}
We note that Souper has been primarily designed with the LLVM IR in mind and requires a well-formed SSA representation of the program under superoptimization.
The biggest challenge with WebAssembly is that there no complete transformation from WASM to SSA. In our pipeline, we work around this by assuming we have access to source code, this alternative path may be valid for plugging other binary format into Souper.

\section{Experiments}

To study the effects and feasibility of applying superoptimization to WASM code, we run the superoptimization pipeline on a benchmark of programs.

\subsection{Benchmark}
The benchmark is based on the Rosetta Code corpus\footnote{\url{http://rosettacode.org}}.
We have selected 12 C language programs that compile to WASM.
Our selection of the programs is based on the following criteria:
\begin{compactenum}[(1)]
\item The programs can be successfully compiled to LLVM IR.
\item They are diverse in terms of application domain.
\item The programs are small to medium sized: between 15 and 200 lines of C code each.
\item They have no dependencies to external libraries.
\end{compactenum}
The code of each program is available as part of our experimental package\footnote{\url{https://github.com/KTH/slumps/tree/master/utils/pipeline/benchmark4pipeline_c}}.

%In addition, our benchmarks include 2 bigger, well-known programs from recent work \cite{Zakai2019}: SQLite (database) and the Lua interpreter. These programs represent real-world codebases with 139765, and 2080 lines of C code, respectively. 

\subsection{Methodology}
\label{method}
To evaluate our superoptimization pipeline, we run it on each program with four Souper configurations:
%1) Inferring only replacements for constant values, 2) Inferring instructions with no more than 2 sentences, i.e. a new replacement is composed by no more than two instructions, 3) CEGIS (Counter Example Guided Inductive Synthesis) for constant values, and 4) enumerative synthesis with no replacement size limit.
\begin{compactenum}[(1)]
\item Inferring only replacements for constant values
\item Inferring replacements with no more than 2 instructions, i.e. a new replacement is composed by no more than two instructions
\item CEGIS (Counter Example Guided Inductive Synthesis, algorithm developed by Gulwani et al. \cite{CEGIS})
\item Enumerative synthesis with no replacement size limit
\end{compactenum}
In the rest of the paper, we report on the best configuration per program. Our appendix website contains the results for all configurations and all programs.

% correctness
With respect to correctness, we rely on Souper's verification to check that every replacement on each program is correct. That means that the superoptimized programs are semantically equivalent. Every candidate search is done with a 300 seconds timeout. For each program, we report the best optimized case over all mentioned configurations. To discuss the results, we report the relative instruction count before and after superoptimization. 

For the baseline program, we ask LLVM to generate WASM programs based on the `wasm32-unknown-unknown' target with the -O3 optimization level.
Our experiments run on an Azure machine with 8 cores (16 virtual CPUs) at 3.20GHz and 64GB of RAM.

\subsection{Results} 
 
\begin{figure}
  \includegraphics[width=\linewidth]{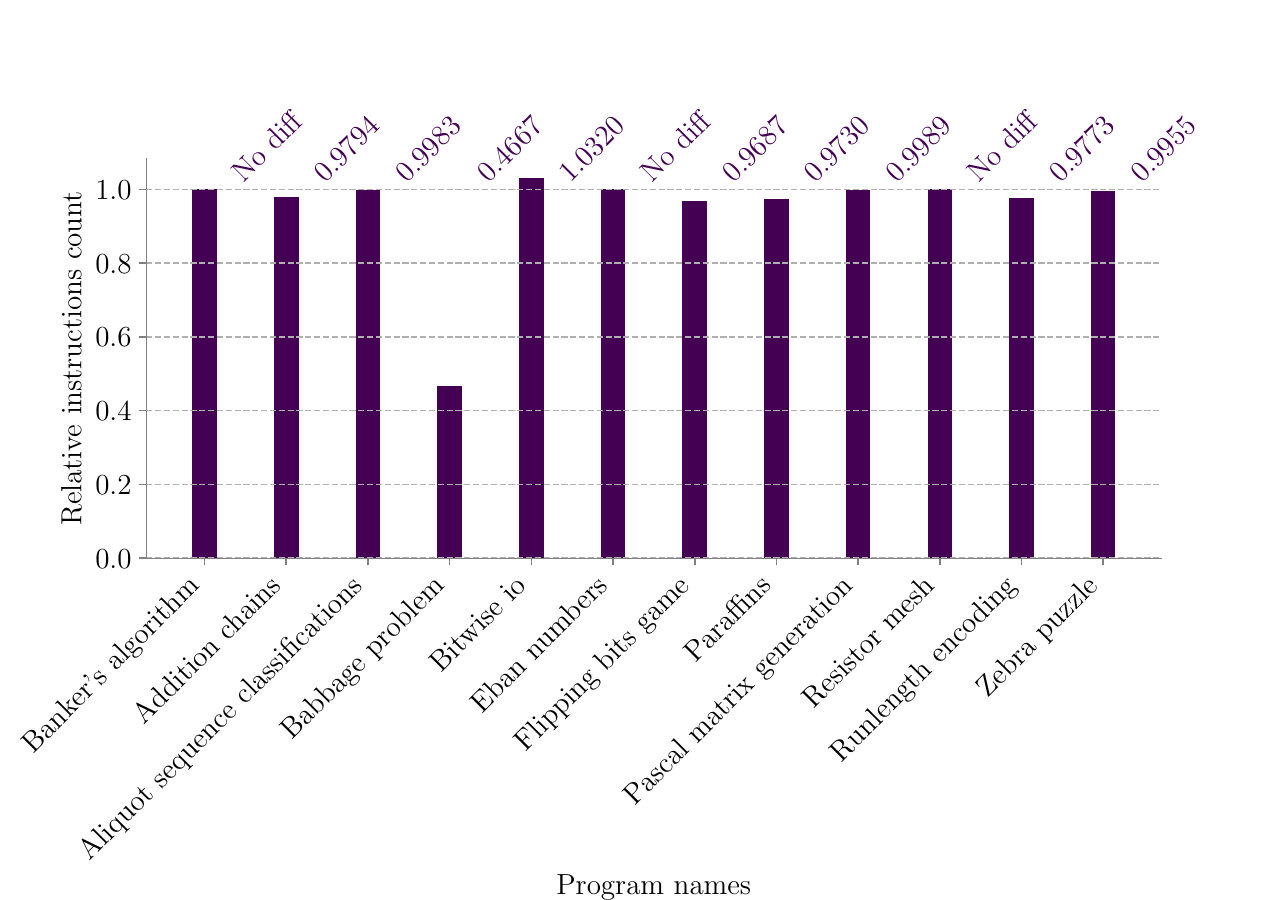}
   \caption{Vertical bars show the relative binary size in \# of instructions. This captures the size difference between the original wasm bytecode and the superoptimized one. The smaller, the better.}
   \label{sizes_fig}
\end{figure}

\autoref{sizes_fig} shows the relative size improvement with superoptimization.

The median size reduction is 0.33\% of the original instruction count over the tested programs. From the 12 tested programs, 8 have been improved using our pipeline whereas 3 have no changes and 1 is bigger (\textbf{Bitwise IO}).
The most superoptimized program is \textbf{Babbage problem}, for which the resulting code after superoptimization is 46.67\% smaller than the baseline version.

\begin{figure}
  \includegraphics[width=\linewidth]{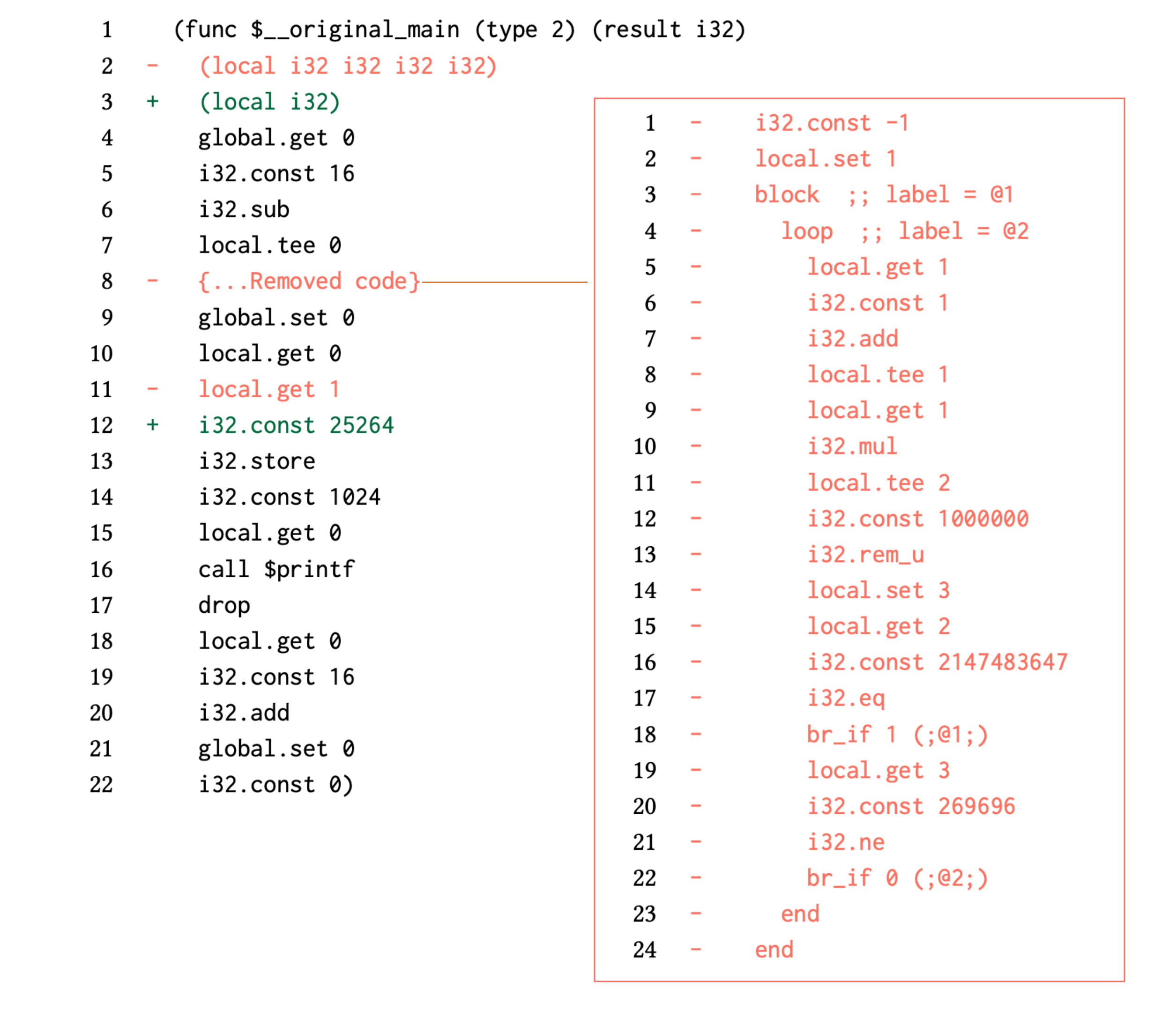}
   \caption{Output of superoptimization WASM bytecode for the Babbage problem program.}
   \label{IR}
\end{figure}

We now discuss the \textbf{Babbage problem} program, originally written in 15 lines of C code\footnote{\url{http://www.rosettacode.org/wiki/Babbage_problem\#C}}. The pipeline found 3 successful code replacements for superoptimization out of 7 candidates.
The best superoptimized version contains 21 instructions, which is much less than the original which has 45 instructions. The superoptimization code difference program is shown in \autoref{IR}. 
Our pipeline, using Souper, finds that the loop inside the program can be replaced with a \texttt{\textbf{const}} value in the top of the stack, see lines 8 and 12 in \autoref{IR}. The value, 25264, is the solution to the Babbage problem. In other terms, the superoptimization pipeline has successfully symbolically executed the problem. 

The \textbf{Babbage problem} code  is composed of a loop which stops when it discovers the smaller number that fits with the Babbage condition below.
\vspace{-0.1cm}
\begin{center}
\begin{tabular}{c}
\begin{lstlisting}[language=C++,basicstyle=\small\ttfamily]
while((n * n) % 1000000 != 269696) n++;
\end{lstlisting}
\end{tabular}
\end{center}

% llvm-opt: rool unroll
In theory, this value can also be inferred by unrolling the loop the correct number of times with llvm-opt.
However, llvm-opt cannot unroll a \texttt{\textbf{while}}-loop because the loop count is not known at compile time.
Additionally, this is a specific optimization that does not generalize well when optimizing for code size and requires a significant amount of time per loop.

% Souper
On the other hand, Souper can deal with this case. The variable that fits the Babbage condition is inferred and verified in the SMT solver. Therefore the condition in the loop will always be false, resulting in dead code that can be removed in the final stage that generates WASM from bitcode.

% Reasons behind low level of improvement
%We analyze the results to identify the reasons that can hinder superoptimization. The only undefined behavior modelled by
%Souper is divide-by-zero. The presence of undefined behaviors or poison values in candidates can leverage in a failed inferring.

% case study
In the case of the \textbf{Bitwise IO} program, we observe an increase in the number of instructions after superoptimization. From the original number of 875 instructions, the resulting count after the Souper pass is increased to 903 instructions. In this case, Souper finds 4 successful replacements out of 207 possible ones. Looking at the changes, it turns out that the LLVM IR code costs less than the original following the Souper cost function. However, the WebAssembly LLVM backend (wasm-ld tool) that transforms LLVM to WASM creates a longer WASM version. This a consequence of the discussion on Souper in \autoref{souper}. In practice, it is straightforward to detect and discard those cases.

\subsection{Correctness Checking}
To validate the correctness of the superoptimized program we perform a comparison of the output of the non-superoptimized program and the superoptimized one.
For 7/12 programs, both versions, non-superoptimized and superoptimized, behave equally and return the expected output. 
For 5/12 programs we cannot run them because the code generated for the target WASM architecture lacks required runtime primitives.

% Since we want to use standard library calls defined in the original programs, we ran the LLVM bitcode instead of the WASM binary to check correctness.

\section{Related Work}

Our work spans the areas of compilation, transformation, optimization and web programming. Here we discuss three of the most relevant works that investigate superoptimization and web technologies.

Churchill et al. \cite{churchill_sound_nodate} use STOKE \cite{bansal_automatic_nodate} to superoptimize loops in large programs such as the Google Native Client \cite{noauthor_welcome_nodate}. They use a bounded verifier to make sure that every generated optimization goes through all the checks for semantic equivalence. We apply the concept of superoptimization to the same context, but with a different stack, WebAssembly. Also, our work offloads the problem of semantic checking to an SMT solver, included in the Souper internals. 

% Souper \cite{bansal_automatic_nodate} automatically generates optimization rules by learning the database of thousands to million superoptimizers. Besides, it uses a SMT solver to verfiy the semantic equivalence of the generated candidate optimizations. Slumps takes a similar approach where it uses a SAT solver to verify the equivalence automatically. It uses synthesis rather than enumeration by learning the information of the dataflow to construct the relevant RHSs.

%GreenThumb \cite{phothilimthana_greenthumb:_2016-1} by Phothilimthana et.al gives an extensible approach for providing a fast and efficient search algorithm for superoptimizers to be used in any ISA. Whereas Optgen \cite{franke_optgen:_2015} by Buchwald et.al is also a recent superoptimizer which operates on IR (Intermediate Representation) unlike our work which operates on assembly level language. It also generates optimizations for symbolic constants unlike Souper, and can generate code optimizations upto a certain cost-limit rather than extracting code-sequences.
Emscripten is an open source tool for compiling C/C++ to the Web Context. Emscripten provides both, the WASM program and the JavaScript glue code. It uses LLVM to create WASM but it provides support for faster linking to the object files. Instead of all the IR being compiled by LLVM, the object file is pre-linked with WASM, which is faster. The last version of Emscripten also uses the WASM LLVM backend as the target for the input code. 
                    
%Cheerp \cite{Cheerps2018} is a tool to generate WASM based on LLVM/Clang standard. It compiles a single C++ code base into a combination of WebAssembly and JavaScript. Cheerp leverages the advantages of WebAssembly (native-like speed and size) and of JavaScript (DOM manipulation and WebAPI access) to generate faster code \cite{Cheerp_2.0_release}. In comparison, our work cares more about WebAssembly itself.

To our knowledge, at the time of writing, the closest related work is the ``souperify'' pass of Binaryen \cite{Binaryen}. It is implemented as an additional analysis on top of the existing ones. Compared to our pipeline, Binaryen does not synthesize WASM code from the Souper output.

\section{Conclusion}

We propose a pipeline for superoptimizing WebAssembly. It is a principled integration of two existing tools, LLVM and Souper, that provides equivalent and smaller WASM programs.

We have shown that the superoptimization pipeline works on a benchmark of 12 WASM programs. As for other binary formats, superoptimization of WebAssembly can be seen as complementary to standard optimization techniques. Our future work will focus on extending the pipeline to source languages that are not handled, such as TypeScript and WebAssembly itself.

\section{Acknowledgements}
This work has been partially supported by WASP program and by the TrustFull project financed by SSF. We thank John Regehr and the Souper team for their support.

\bibliographystyle{ACM-Reference-Format}

%\balance
\bibliography{main}

\newpage

\end{document}